\documentclass[a4paper,11pt]{article}
\pdfoutput=1 
\usepackage{jheppub} 
\usepackage{fancyhdr}
\usepackage{graphicx}
\usepackage{xspace}
\usepackage{rotating}

\usepackage{braket}

\DeclareSymbolFont{extraup}{U}{zavm}{m}{n}
\DeclareMathSymbol{\varheart}{\mathalpha}{extraup}{86}
\DeclareMathSymbol{\vardiamond}{\mathalpha}{extraup}{87}

\usepackage{amsfonts}
\usepackage{pifont}

\newcommand{\be}{\begin{equation}}
\newcommand{\ee}{\end{equation}}
\newcommand{\bea}{\begin{eqnarray}}

\newcommand{\eea}{\end{eqnarray}}

\newcommand{\Rmnum}[1]{\expandafter\@slowromancap\romannumeral #1@}

\def\mev{\,{\rm MeV}}
\def\gev{\,{\rm GeV}}

\begin{document} 

 \title{\boldmath Enabling Electroweak Baryogenesis through Dark Matter}
 
\author[1,2]{Marek  Lewicki}
\author[2,3]{Tanja Rindler-Daller}
\author[2]{James D. Wells}

\affiliation[1]{
Institute of Theoretical Physics, Faculty of Physics, University of Warsaw ul. Pasteura 5, 02-093 Warsaw, Poland}
\affiliation[2]{Department of Physics and
Michigan Center for Theoretical Physics, University of Michigan, 450 Church St., Ann Arbor, MI 48109, USA} 
\affiliation[3]{
Institut f\"ur Astrophysik, Universit\"atssternwarte Wien, University of Vienna, T\"urkenschanzstr.17, 1180 Vienna, Austria}
\emailAdd{marek.lewicki@fuw.edu.pl}
\emailAdd{daller@umich.edu}
\emailAdd{jwells@umich.edu}


\abstract{We study the impact on electroweak baryogenesis from a swifter cosmological expansion induced by dark matter. 
We detail the experimental bounds that one can place on models that realize it, and we investigate the modifications of 
these bounds that result from a non-standard cosmological history.
The modifications can be sizeable if the expansion rate of the Universe increases by several orders of magnitude.
We illustrate the impact through the example of scalar field dark matter, which can alter the cosmological history enough to enable a strong-enough first-order phase transition in the Standard Model  
when it is supplemented by a dimension six operator directly modifying the Higgs boson potential.
We show that due to the modified cosmological history, electroweak baryogenesis can be realized, while keeping deviations of the triple Higgs coupling below HL-LHC sensitivies. The required scale of new physics to effectuate a strong-enough first order phase transition can change by as much as twenty percent as the expansion rate increases by six orders of magnitude. 
} 


\maketitle
\flushbottom


\section{Introduction} \label{sec:intro}

The discovery of the Higgs boson with mass $m_h \simeq 125$ GeV at the Large Hadron Collider (LHC)
\cite{Aad:2012tfa,Chatrchyan:2012xdj} confirmed that electroweak symmetry is broken due to the vacuum expectation value (vev) 
of a scalar field. However, the symmetry was restored in the early Universe due to high temperature modifications
of the Higgs boson properties.
The dynamics of a phase transition from a symmetric phase in the hot early Universe to the present-day
broken phase at low temperature is described by finite-temperature field theory. While the high-temperature Higgs dynamics is not 
directly measurable at a collider, it is tightly related to the currently probed zero-temperature potential.

In this paper, we wish to study a baryogenesis scenario \cite{Kuzmin:1985mm,Cohen:1993nk,Riotto:1999yt,Morrissey:2012db}
in which the observed baryon asymmetry of the Universe is created during the electroweak phase transition (EWPT).
This requires baryon number violation, both $C$ and $CP$ violation and the departure from thermal equilibrium \cite{Sakharov:1967dj}.
The last condition can be fulfilled if the phase transition is first order.    
However, in the Standard Model (SM) it is second order for a Higgs mass of $125$ GeV, and the field transitions smoothly into its new non-symmetric minimum which develops as the temperature drops.
Thus, models of electroweak baryogenesis require new physics near the electroweak scale in order 
to generate a barrier between the symmetric phase and the broken phase \cite{Arnold:1992rz,Kajantie:1996qd}.
Such models gained renewed attention recently, as the experimental accuracy with which we know the Higgs properties increases and models predicting modification to its potential can be probed \cite{Cohen:2012zza,Katz:2014bha}. 

In this paper, we will discuss a generic model of new physics studied previously in \cite{Grojean:2004xa,Delaunay:2007wb}.
The Higgs potential is modified by a non-renormalizable dimension six operator, 
which is suppressed by a cutoff scale $\Lambda$. In the references above it was shown that this model can facilitate a first-order PT, depending on the value of $\Lambda$ and the Higgs boson mass.
Now that the Higgs mass has been determined, the only remaining free parameter is the cutoff scale $\Lambda$.
Also in the references above, the usual assumption that the EWPT proceeds in a radiation-dominated Universe was applied. In this work our main focus will be to determine the consequences of relaxing this assumption.
We emphasize that we will not discuss the production of baryons during the EWPT, but rather the necessary condition for the baryon asymmetry to avoid being ``washed out'' after the phase transition. That is, our analysis concerns the decoupling of fermion number violating processes, due to a modified expansion history.

In the standard cosmological model, the Universe is radiation-dominated from the end of reheating to the time of
matter-radiation equality around $400,000$ yrs after the Big Bang. During this time, a plethora of phase transitions occurred,
among them the EWPT at a scale around $T \sim 100$\gev. While the good agreement between Big Bang nucleosynthesis (BBN) models and measurements of the primordial elemental abundances imply that the Universe was radiation-dominated during and after BBN (i.e., after
a time corresponding to $T \simeq 1\,$MeV), the expansion history before BBN is still very poorly constrained. 
This is equivalent to the statement that the energy density in the early Universe could have been distributed among components in such a way that some of them may have sufficiently decayed or transformed into radiation, and so their presence may not show up through measurements
of the energy density at later epochs.
Since extensions of the SM predict new particles and energy constituents, the question of how those will impact the early Universe arises naturally.

The difficulty for electroweak baryogenesis (EWBG) that we will discuss arises when the Universe returns to thermal equilibrium after the phase transition. Then, the same sphaleron processes that could have created the baryon asymmetry during the transition can wash it away, if their damping in the broken phase is not sufficient. One way to avoid this problem is to generate a large potential barrier, such that these processes are damped enough after the transition. However, cosmological freeze-out due to a fast expansion of the Universe works in the same direction, see \cite{Joyce:1996cp,Joyce:1997fc}.
In order to obtain a higher expansion rate of the Universe than in the standard case, we require that the dominant energy density during EWBG decreases faster than radiation which has to dominate the Universe later during BBN.
From the point of view of baryogenesis we can remain agnostic to what cosmological model modifies the evolution of the Universe during this early epoch. Indeed, our results are applicable to a large class of cosmological models and do not depend on the detailed implementation of such models.
However, it is still an important question what agent could give rise to such a modification.
It turns out that there exist models in which that agent is actually the cosmological dark matter (DM). Hence, in these scenarios
it would be the DM itself which facilitates EWBG, lending further motivation to such DM models.
Specifically, we will discuss one of those models, namely scalar field dark matter (SFDM), in which the entire dark matter abundance is described as a complex scalar field,
see \cite{Li:2013nal}. The underlying particles are ultra-light bosons, which condensed into their ground state soon after their birth.
As a result, DM can thereafter be described as a classical field with a conserved $U(1)$-charge, which is effectively the conserved DM abundance, which is chosen to match the present-day value. 

Thus, our approach is to combine a simple and straightforward theory of first-order electroweak symmetry breaking phase transition ($|H|^6$ theory) with a simple and straightfoward modification of the cosmological history (with the specific example of SFDM theory which also explains the dark matter abundance) to suppress sphaleron rates, thereby enabling the possibility of electroweak baryogenesis.
The outline of this paper is as follows.
We begin by describing the $\phi^6$ model, relevant experimental bounds, and the dark matter in Section \ref{sec:part_model}. 
In Section \ref{sec:EWPTdynamics}, we compute the temperature at which the EWPT takes place, and contrast that to the 
critical temperature 
often used in the literature.
This also allows us to properly include corrections to this temperature, coming from a modification of the expansion history.
We show that this correction is very small and can be safely neglected, as expected.  In Section~\ref{sec:freezout}, we describe the 
modification of Standard Model $SU(2)$ sphalerons responsible for generating the  baryon asymmetry during the 
phase transition, and its ``wash-out'' after the phase transition has completed. 
Here is where the main modification from the increased expansion rate of the Universes comes in by making the sphalerons decouple faster, after the
phase transition has completed.
This results in an increased amount of remaining baryons, which, in turn, increases the minimal energy scale $\Lambda$ of new physics required for baryon asymmetry preservation. We find that this change can be significant, moving $\Lambda$ by twenty percent as the Hubble parameter increases by six orders of magnitudes compared to the one in the standard case.
While the modification does not seem huge, it actually means 
circumventing the sphaleron bound altogether, since it brings us very close to the cutoff values required for a first-order phase transition to begin with.  
On the particle experimental side, it means that, with the assistance of SFDM, say, our model can predict a modification 
of the Higgs self-coupling which is only slightly bigger than $1\sigma$ away from the corresponding SM central value. This result will
not even be challenged by the tighter bounds provided by the high luminosity stage of the LHC (HL-LHC).
 
\section{The Particle Model}\label{sec:part_model}

In order to effectuate a first-order phase transition of electroweak symmetry breaking (EWSB), we need a particle model that goes beyond the SM.
There are numerous ideas in the literature that accomplish this.
Perhaps the simplest idea is to not introduce new propagating degrees of freedom, but to merely add a single higher-dimensional operator $|H|^6$ which can create a potential barrier between two local minima at the critical temperature, thereby achieving the first-order phase transition as the temperature drops. 
This will be the approach we pursue in this paper, and in the next subsection we review the details of this model and also the finite temperature field theory formalism needed to investigate the phase transition.

Achieving a first-order phase transition is not enough; it must be strong enough. One can either scrap the simple EWSB $|H|^6$ and pursue richer variants, or one can consider if the Universe has a non-standard cosmological evolution that redefines what is the acceptable strength of the first-order phase transition for sphalerons to not wash out any baryon asymmetry that might have been created during the transition. It is this approach that we take.  We can be agnostic as to what kinds of dynamics enable a different cosmological evolution during the early time of the EWPT and parametrize simply what is needed. We will express that attitude at times and show results that are generally applicable to one's favorite theory. However, to be concrete, we will present results within the specific framework of SFDM, which provides an excellent and motivated illustration of how the dark matter background can affect the viability of the first-order phase transition. To that end, we provide in this section some 
additional descriptions of the SFDM model and comments on how it affects the cosmological evolution. We also mention in passing other models with a similar effect in the early Universe.

\subsection{The $\phi^6$ EWSB Theory}\label{sec:particlemodel}

Here we describe the particle physics dynamics of having an additional $|H|^6$ term in the EWSB Higgs potential. The new non-renormalizable coupling is suppressed by a certain mass scale $\Lambda$. 
Above that scale, new degrees of freedom become fully dynamical, and the underlying particle model cannot be described in the language of our effective theory.
Restricting ourselves to processes around the electroweak scale,
we will consider the following potential 
\begin{equation}\label{eqn:classpot}
V(H)=-m^2|H|^2+\lambda|H|^4+\frac{1}{\Lambda^2} |H|^6,
\end{equation}
with
$H^T=\left(\chi_1+i\chi_2,\varphi+i\chi_3\right)/\sqrt{2}$.
We assume only the real part of the neutral component has a vev: $\varphi=\phi+v$. The physical Higgs boson is $\phi$, 
which leads to the following tree level potential
\begin{equation}\label{eqn:classpottree}
V(\phi)^{\textrm tree}=-\frac{m^2}{2}\phi^2+\frac{\lambda}{4}\phi^4+\frac{1}{8}\frac{\phi^6}{ \Lambda^2}.
\end{equation} 
The field-dependent masses take the form
\begin{equation}
\begin{split}
m^2_h(\phi) & = -m^2+3\lambda \phi^2 +  \frac{15}{4}\frac{\phi^4}{\Lambda^2},  \\
m^2_{\chi_i}(\phi)& =  -m^2+\lambda \phi^2 + \frac{3}{4}\frac{\phi^4}{\Lambda^2},\\
m^2_W(\phi)& = \frac{g^2}{4}\phi^2,
\quad m^2_Z(\phi)=\frac{g^2+g'^2}{4}\phi^2, 
\quad\ m^2_t(\phi)=\frac{y_t^2}{2}\phi^2,
\label{eqn:massh}
\end{split}
\end{equation}
where $g, g'$ and $y_t$ are the gauge boson and Yukawa couplings, respectively.

\begin{figure}[t]
\begin{minipage}[t]{0.45\linewidth} 
\includegraphics[height=6.7cm]{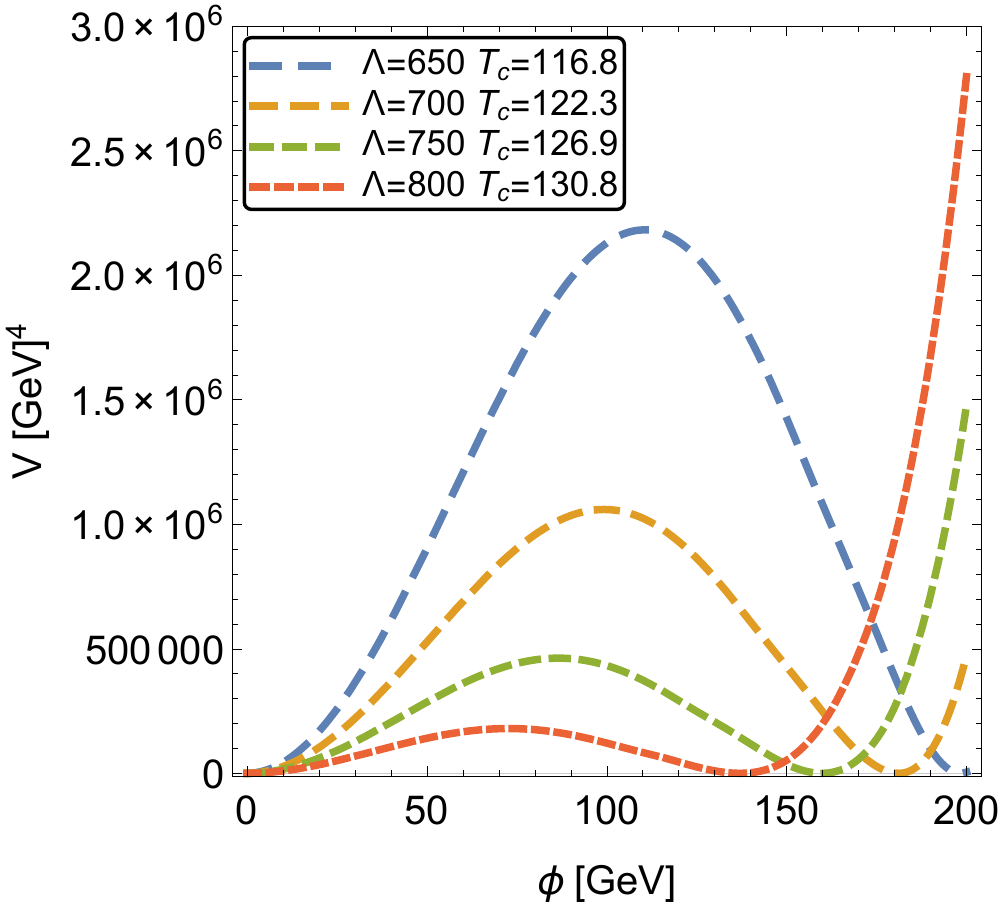} 
\end{minipage}
\begin{minipage}[t]{0.45\linewidth}
\includegraphics[height=6.7cm]{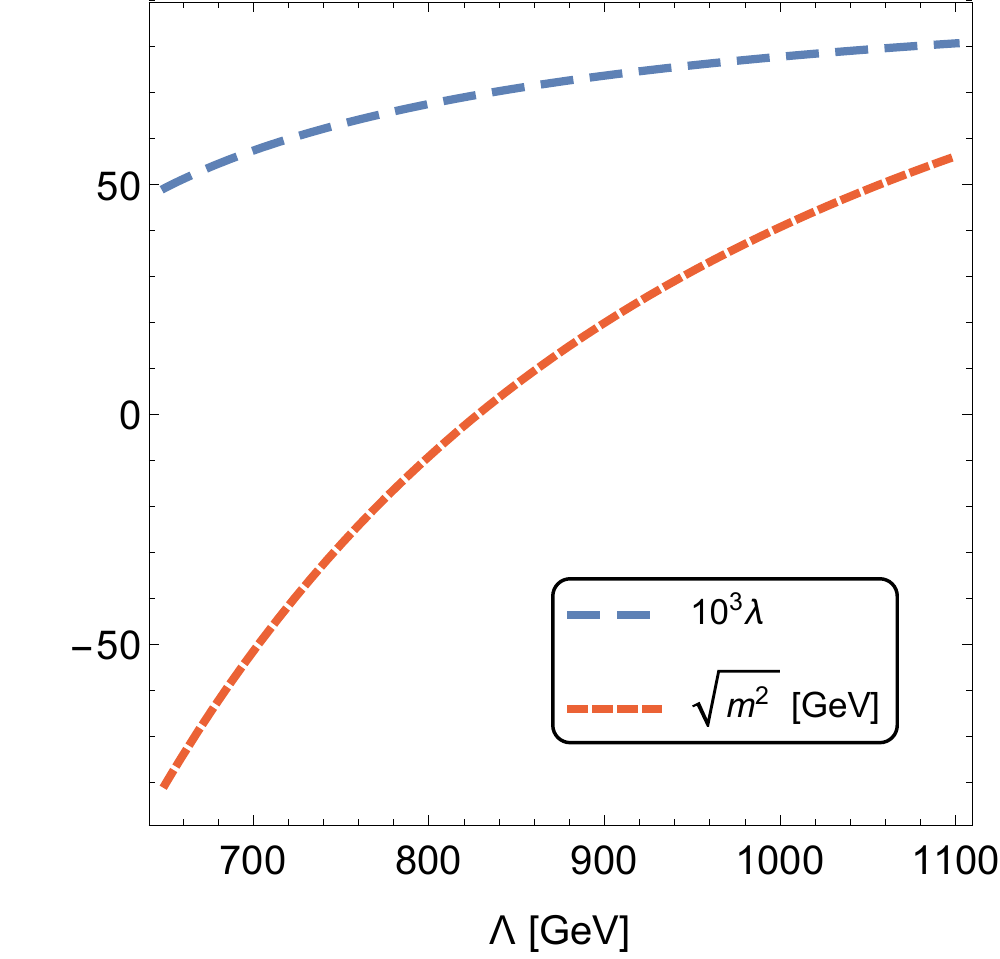}  
\end{minipage}
\caption{
The left panel shows examples of potentials at their critical temperatures for several values of $\Lambda$.
The right panel depicts the values of $m$ and $\lambda$ as functions of $\Lambda$ (all values except for the dimensionless 
$\lambda$ are expressed in \gev).}
\label{potplot}
\end{figure}

Following the prescription from \cite{Delaunay:2007wb}, where a very similar potential was considered,
we include thermal and loop corrections as follows,
\begin{equation}\label{eqn:Veff}
\begin{split}
V_{eff}(\phi,T)&=-\frac{m^2}{2}\phi^2+\frac{\lambda}{4}\phi^4
           +\frac{1}{8}\frac{\phi^6}{\Lambda^2} + \sum_{i=h,\chi,W,Z,t}n_i\frac{m_{i}^4(\phi)}{64\pi^2}\left[\log\frac{m^2_{i}(\phi)}{\mu^2}-C_i\right] 
            \\
 &  +\sum_{i= h,\chi,W,Z}\frac{n_iT^4}{2\pi^2} J_b\left(\frac{m^2_i(\phi)}{T^2}\right)+\sum_{i= t}\frac{n_iT^4}{2\pi^2} J_f\left(\frac{m^2_i(\phi)}{T^2}\right) 
  \\
 &  +\sum_{i=h,\chi,W,Z,\gamma}\frac{\bar{n}_iT}{12\pi}\left[m^3_i(\phi)-\left(m^2_i(\phi)+\Pi_i(T)\right)^{3/2}\right].
\end{split}
\end{equation}
The coefficients read $n_{\{h,\chi,W,Z,t\}}=\{1,3,6,3,-12\}$, $\bar{n}_{\{h,\chi,W,Z,t\}}=\{1,3,2,1,1\}$, $C_i=3/2$ for $i=h,\chi,t$ and $C_i=5/6$ for $i=W,Z$,
the functions $J$ are given by 
\begin{equation}
J_{b/f }\left(\frac{m^2_i(\phi)}{T^2}\right)=\int_0^\infty dk \, 
k^2\log\left[1\mp {\rm exp}\left(-\sqrt{\frac{k^2+m_i^2(\phi)}{T^2}} \right) \right].
\end{equation}
The mass corrections $\Pi_i$ in \eqref{eqn:Veff} result from the {\it ring-improvement}
of the finite temperature potential, which is a resummation of the so-called {\it daisy} diagrams
that become enhanced at high temperature in the limit of zero boson mass. 
In our model, these mass shifts read \cite{Carrington:1991hz,Delaunay:2007wb}
\begin{equation}
\begin{split}
\Pi_{h ,{\chi_i}}(T)&=\frac{T^2}{4v_0^2}\left(m_h^2+2m_W^2+m_Z^2+2m_t^2\right)-\frac{3}{4} T^2 \frac{v_0^2}{\Lambda^2}
\\
\Pi_W(T)&=\frac{22}{3}\frac{m^2_W}{v_0^2}T^2\\ 
\end{split}
\end{equation}
and the shifted masses of $Z$ and $\gamma$ ($m^2_{Z/\gamma}+\Pi_{Z/\gamma}(T)$) are eigenvalues of the following mass matrix, including thermal corrections
\begin{equation}
\begin{pmatrix}
\frac{1}{4} g^2 \phi^2+\frac{11}{6}g^2T^2 & -\frac{1}{4}g'^2 g^2 \phi^2 \\
-\frac{1}{4}g'^2 g^2 \phi^2 & \frac{1}{4} g'^2 \phi^2+\frac{11}{6}g'^2T^2
\end{pmatrix}.
\end{equation}
The values of the parameters $\lambda$ and $m$ are calculated from the conditions that 
\begin{equation}\label{masseq}
\begin{split} 
V'_{eff}(\phi,T=0)|_{\phi=v_0}=0, \quad V''_{eff}(\phi,T=0)|_{\phi=v_0}=m_h,
\end{split}
\end{equation}
i.e., requiring the observed masses of the Higgs boson $m_h=125.09$\gev, as well as those for the gauge bosons via the Higgs ground state of 
$v_0 := \langle \phi (T=0) \rangle = 246.2$\gev.
The resulting values of the parameters $m$ and $\lambda$, as well as examples of potentials, are shown in Figure~\ref{potplot}.

Using higher order corrections to the Higgs mass would result in a mass parameter higher by a few percent and $\lambda$ smaller by a few percent  \cite{Buttazzo:2013uya}. This results in a slightly bigger barrier and stronger phase transition; however, it is negligible  compared to the modification coming from the non-renormalizable correction.
It is also known that the two-loop thermal potential predicts a bigger thermal barrier between the vacua and therefore results in a stronger phase transition \cite{Arnold:1992rz,Arnold:1993bq}. However, our aim is to illustrate the effects of modified cosmology and a stronger phase transition would only serve to strengthen our conclusions.

We will limit our considerations to cutoff scales smaller than $\Lambda \approx 1100$ GeV. Above that scale, the phase transition is as weak as in the Standard Model (SM) with $m_h\approx 80$\gev, where the barrier between vacua is actually negligible and lattice simulations show results similar to a second-order phase transition \cite{Kajantie:1996mn}.
In that case, even if the sphalerons can be decoupled during the phase transition, no asymmetry will be created to begin with, so the model would already be ruled out.

\subsection{The Triple Higgs Coupling}\label{sec:expexcl}

While all UV complete models realizing EWBG predict various modifications 
to Higgs properties, our effective theory differs from the SM only in terms of the Higgs potential. 
All other modifications of Higgs properties in the language of an effective theory are simply unrelated, until one chooses a specific UV completion. Consequently, we will not discuss them.

In this approach, the only directly affected measurable Higgs property is the triple-Higgs coupling related to the third derivative 
of the zero-temperature potential \eqref{eqn:Veff},
\begin{equation}
\lambda_3 = \frac{1}{6} \left. \frac{d^3 V_{eff}(\phi,{T=0})}{d \phi^3} \right|_{\phi=v_0}.
\end{equation}
This coupling can be measured in double Higgs production events; however, the cross-section for producing a single Higgs boson
is roughly three orders of magnitudes larger.
This is why high-luminosity experiments are required for a reliable measurement.
LHC in its high-luminosity phase (HL-LHC) will be able to determine the value of $\lambda_3$ with roughly $40\%$ accuracy \cite{ATLAS-Collaboration:2012jwa,Goertz:2013kp,Barger:2013jfa,Barr:2014sga}.  

\begin{figure}[t] 
\begin{center}
\includegraphics[height=6.5cm]{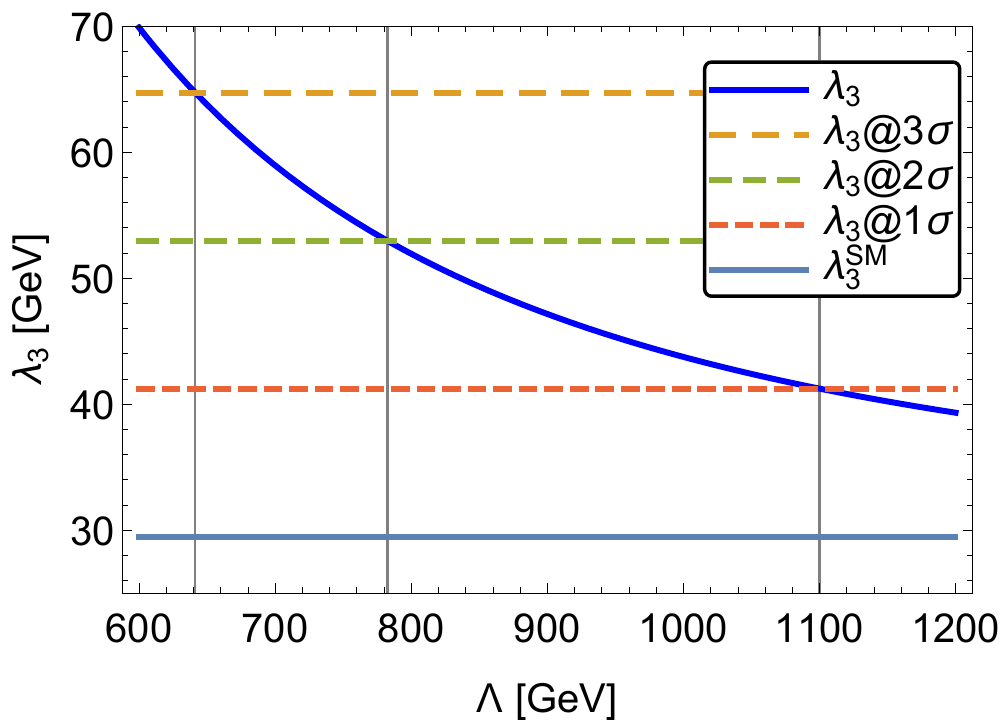} 
\end{center}
\caption{
Values of the triple Higgs coupling $\lambda_3$ as a function of the cutoff scale (dark blue line), along with the SM value (light blue) and HL-LHC experimental sensitivity at $1$, $2$ and $3 \sigma$ (dashed lines). 
The thin vertical lines point to cutoff values corresponding to these bounds which are $\Lambda \approx 1102, 783$ and $641$ GeV, respectively.
\label{lambda3mplot}
}
\end{figure}

Figure~\ref{lambda3mplot} shows the value of $\lambda_3$ in our model as a function of the cutoff scale $\Lambda$,
along with the SM value and the HL-LHC experimental sensitivity at $1$, $2$ and $3 \sigma$, respectively. 
The smaller the cutoff scale (i.e., the larger the deviation from the SM), the larger the coupling $\lambda_3$.
This allows us to explicitly calculate the reach of HL-LHC through $\lambda_3$ measurements, 
in terms of the cutoff scale of new physics. The resulting scales are $\Lambda \approx 1102, 783$ and $641$ GeV,
corresponding to $1$, $2$ and $3 \sigma$ deviations in the measurement, respectively. 

 \subsection{Dark Matter as a Stiff Fluid}\label{sec:sfdm background}
 
As discussed in the Introduction, the new energy component must decay faster than radiation in order to make a cure to the problem of EWBG. Equivalently, the 
equation-of-state (EOS) of that component must be stiffer than that of radiation, i.e., $p = w \rho$ with $w > 1/3$. The gain is maximal for a ``stiff EOS'' with $w=1$ (still 
consistent with causality). The nature of this component could arise from diverse origins. Previously, the agent in \cite{Joyce:1996cp, Joyce:1997fc} has been thought to be a 
(real) relic scalar field, either different from the inflaton, or the inflaton itself which transitions into the required stiff phase. Indeed, such a phase can easily arise,
if the kinetic mode of a scalar field dominates its evolution (termed ``kination'' in \cite{Joyce:1996cp}).

Our specific example affecting EWSB of the $|H|^6$ potential is scalar field dark matter (SFDM), in which the entire cosmological DM is described as a \textit{complex} scalar field. 
Its cosmological evolution has been studied first analytically in \cite{ALS2002}, but more accurately in 
\cite{Li:2013nal}. The adopted Lagrangian is of a very generic form,
\begin{equation} \label{Lag}
\mathcal{L} = \frac{\hbar^2}{2m_s}g^{\mu \nu}\partial_{\mu}\psi^*\partial_{\nu}\psi - \frac{1}{2}m_s c^2|\psi|^2 - 
\frac{\lambda_s}{2}|\psi|^4.
\end{equation}
$m_s$ is the DM boson mass and the energy-independent
boson coupling strength is chosen to be repulsive or zero, $\lambda_s \geq 0$. The SFDM model parameters, $m_s$ and $\lambda_s$, 
need to be tiny for reasons of DM structure formation, see e.g. \cite{ALS2003, Rindler-Daller:2013zxa}. 
Indeed, one major motivation to study SFDM is its potential to resolve certain small-scale problems faced by
standard collisionless cold dark matter (CDM), in that it provides a characteristic (Jeans) scale, prohibiting gravitational collapse below that scale, as follows.

If CDM is due to weakly interacting massive particles (WIMPs), the predicted minimum clustering scale 
allows substructure down to the order of earth-mass microhalos with radius of order the solar system. As a result, CDM predicts not only
very high DM densities in the centers of galaxies, but also a much larger number of satellite galaxies around hosts like the Milky Way.
Both predictions have been continuously challenged by observations of DM-dominated galaxies, which report deviations up
to scales of one kpc (see e.g. \cite{WD2013CDMP} for a review). This discrepancy is one major reason to study alternatives to the CDM/WIMP paradigm. The Jeans scale of SFDM can be of order kpc,
if the mass $m_s \simeq 10^{-22}$ eV. Higher masses are possible, if a positive coupling strength is included; 
fiducial values of $\lambda_s \approx 10 ^{-62}$ eV cm$^3$ correspond to dimensionless couplings of order $\tilde \lambda_s \approx 10^{-92}$. 
These values are many orders of magnitudes below typical values for the QCD axion for which $m_s \simeq 10^{-5}$ eV and $\tilde \lambda_s \approx 10^{-53}$.

These values we use for the SFDM parameters are purely phenomenological and we do not address possible fine-tuning issues, which we believe are not any worse than the large hierarchy problem in the SM. 

While the mass term in (\ref{Lag}) will ensure that SFDM behaves like CDM with $p \simeq 0$ in the late Universe, the presence of the other terms
will render SFDM a relativistic species in the early Universe. When the quartic term dominates (assuming $\lambda_s > 0$), SFDM behaves radiation-like,
i.e., $p \simeq \rho/3$. However, it is radiation that will dominate the cosmic energy budget in that phase. Finally, even earlier in the evolution,
the kinetic term in (\ref{Lag}) will give rise to a stiff EOS of SFDM, $p \simeq \rho$, in which SFDM will dominate over all the other cosmic components.
Note that SFDM transitions \textit{dynamically} through all its phases, and that the stiff phase arises in all models, with or without self-interaction \cite{Li:2013nal}.
While the presence of a stiff epoch for real scalar fields requires the choice of an appropriate potential, it is a generic feature of complex SFDM.
It is the angular part due to the phase of the complex field in the kinetic energy term in (\ref{Lag}) that dominates for small scale factor, whereas this term
is absent in the real field case.

The requirement of compliance with cosmological observables, 
which probe the relativistic degrees of freedom during the cosmic history of our Universe,
determines the allowed parameter space of SFDM models. The observables are notably BBN and the time of matter-radiation equality. 
We refer the reader to \cite{Li:2013nal}, in which a detailed analysis has been presented. An additional important observable of future relevance is given by primordial gravity waves
from inflation, the impact of which on SFDM will be published elsewhere \cite{LRS2:2014}.
Another issue raised recently, and not discussed up to now in the context of SFDM, are the isocurvature constraints which could call into question the identification of SFDM as dark matter. However, a naive reinterpretation of bounds recently derived in \cite{Kainulainen:2016vzv} for a very weakly coupled scalar suggests these bounds do not exclude SFDM. Nevertheless this is a very interesting issue which should be investigated further, but whose results would not affect our conclusions since it could only exclude part of our parameter space at best, and not all of it.
 To our amazement, we found that SFDM models do exist which
fulfill all the constraints and provide a high enough expansion rate during EWBG. A fiducial case has been presented in \cite{Rindler-Daller:2015lua}.

We want to stress, however, that the stiff epoch transitions rapidly into the radiation-dominated epoch due to the high power-law decay, no
matter whether the underlying agent is some generic scalar field that simply keeps diluting away faster than radiation, or
whether it is SFDM that dynamically transitions from a stiff phase to a radiation-like phase. Therefore, the analysis of our paper 
does not depend on a model specification for the stiff epoch, and applies to models different from SFDM, as well. However, as a matter of fact,
 it is a very attractive feature 
of SFDM that it could not only explain the dark matter in the Universe, but potentially also enable EWBG. 

Indeed, to lend further motivation to our analysis, we shall mention that many more models do exist, which could give rise to an early stiff phase. 
Anisotropic cosmologies have been known to include a stiff epoch, and implications have been studied in \cite{KT1990}. 
A stiff phase may be also due to moduli fields, as discussed in \cite{Giovannini}, and which occur frequently in extensions to the SM. On the other hand, a DM model with a stiff phase, which is completely different from SFDM, has been studied in
\cite{SBB}. There, the authors consider warm, self-interacting DM (WSIDM) with or without a collisionless CDM component added to it. The small-scale problems mentioned above
are a prime motivation here, too, to study this DM candidate. The typical mass range of WSIDM particles is 1-10 keV and the DM self-interaction is mediated by vector mesons.
It turns out that the energy density due to DM self-interactions decays as $a^{-6}$, hence dominates in the early Universe. As with SFDM, that contribution has to be constrained by BBN, see \cite{SBB}. In fact, while all such models need to be constrained by BBN and possibly primordial gravitational waves, they may actually help to facilitate EWBG, if the
corresponding stiff phase is allowed to last during the electroweak phase transition. We believe that this feature provides additional motivation for such models.

\section{Electroweak Phase Transition}\label{sec:EWPTdynamics}

Below the critical temperature $T_c$, the minimum that breaks electroweak symmetry becomes
the global minimum of the potential, while the field is still in the symmetric 
local minimum because the two minima are separated by a potential barrier generated 
due to thermal fluctuations, as described in the previous section.
The transition proceeds via thermal tunnelling, 
which can be described in terms of the spontaneous nucleation of bubbles of the broken phase (with non-zero vev) 
in the symmetric background. After nucleation the bubbles grow, converting false vacuum into true one, until the whole 
Universe transitions into the broken phase.  

The crucial value for finding the temperature of the phase transition is the probability of nucleation of a bubble. This probability 
per volume $\mathcal{V}$ is given by \cite{Linde:1981zj, Linde:1980tt}
\begin{equation}\label{eq:decaywidth}
\Gamma/\mathcal{V} \approx T^4 \exp\left(-\frac{S_3(T)}{T}\right),
\end{equation}
where
\begin{equation} \label{actionfunc}
S_3=4\pi \int dr  r^2\left[\frac{1}{2}\left(\frac{d\phi}{dr}\right)^2+V(\phi,T)\right],
\end{equation}
is the action for the solution of the equation of motion that corresponds to the nucleating bubble.
We aim to find the solution with smallest action, i.e., the most symmetric one.  
Assuming an $\mathcal{O}(3)$ symmetric bubble, the equation of motion for the field takes the form
\begin{equation}\label{eqn:scalarEOM}
\frac{d^2 \phi}{dr^2}+\frac{2}{r}\frac{d \phi}{dr}+\frac{\partial V(\phi,T)}{\partial\phi}=0,
\end{equation}
with boundary conditions given by
\begin{equation}
\phi (r\rightarrow\infty)= 0 \mbox{\ \ and\ \ } \frac{d\phi(r=0)}{dr}=0. 
\end{equation}

The question at what temperature the phase transition will proceed and the bubbles 
will percolate also depends on the expansion rate of the Universe. We will assume that 
the phase transition occurs at a temperature $T_n$, at which at least one bubble appears in every horizon.

\subsection{Radiation Domination}
The usual assumption used in the literature is that for $T\approx \mathcal{O}(100 \, {\rm GeV})$ the Universe
is dominated by radiation \cite{Quiros:1999jp},
whose energy density decreases with scale factor $a$ as
\begin{equation}
\tilde{\rho}_{R} =\frac{\rho_{R}}{a^4}.
\end{equation}
Neglecting all the other cosmic components, we can solve the Friedmann equation
\begin{equation}
H^2=\left( \frac{\dot{a}}{a} \right)^2 = \frac{8\pi}{3 M_{p}^2}\frac{\rho_R}{a^4},
\end{equation}
and calculate the volume of the Universe as
\begin{equation}\label{eq:universvolumeRAD}
\mathcal{V}_H(T)=\left(a\int \frac{dt}{a}\right)^3 = 8 \zeta^3 \frac{M_p^3}{T^6},
\end{equation} 
where $\zeta=\frac{1}{4\pi}\sqrt{\frac{45}{\pi g_*}}\approx 2\times 10^{-3}$, assuming the SM number of degrees of freedom
$g_*=106.75$, which is approximately constant in the range of temperatures of interest to us.
Using \eqref{eq:decaywidth}, our condition for one bubble to be nucleated within each horizon then translates to
\begin{equation}\label{eqn:TnucleationRAD}
 \int_{T_n}^{\infty} \Gamma dT=\int_{T_n}^{\infty}\frac{d T}{T}\left( \frac{2\zeta M_p}{T} \right)^4
 \exp\left(-\frac{S_3(T)}{T}\right)=1.
\end{equation}
Before pointing out the importance of using the nucleation temperature $T_n$ instead of the critical temperature $T_c$,
we will discuss how the above result changes due to the modification of our cosmological model.
\subsection{Cosmological Modification}
Now, we want to generalize the foregoing calculation by including a modified expansion epoch before the usual radiation-dominated epoch.
We will assume that the energy density of the new energy constituent redshifts as 
\begin{equation}
\tilde{\rho}_S = \frac{\rho_S}{a^n},
\end{equation}
with $n > 4$. A stiff EOS corresponds to $n=6$. As we move towards earlier times, the contribution of the new component quickly dominates the total energy density.
 Thus, at EWBG we can use a simplified Friedmann equation, including only the new dominating component
\begin{equation}\label{eq:SFDMfriedmann}
H^2=\left( \frac{\dot{a}}{a} \right)^2 = \frac{8\pi}{3 M_{p}^2} \frac{\rho_S}{a^n}.
\end{equation}
We performed calculations in which we include both, radiation and the new component. However, the results can only be expressed using special functions, and we found that the correction coming from neglecting the radiation component
 is completely negligible. Thus, for the sake of clarity we will present only the simplified calculation.
Assuming that the new energy component does not interact with SM degrees of freedom (as is the case for SFDM), we can use the standard 
relationship between temperature and scale factor, namely
\begin{equation}\label{eq:rhorad}
\tilde{\rho}_R=\frac{\rho_{R}}{a^4}=\frac{\pi^2}{30}g_* T^4,
\end{equation} 
which allows us to obtain the scale factor as a function of temperature.
Analogously to the previous subsection, we now calculate the horizon volume
\begin{equation}
\mathcal{V}_H(T)=\left(a\int \frac{dt}{a}\right)^3 =\frac{M_p^3 2^{\frac{3}{8} (5 n-4)} \left(\frac{\pi }{3}\right)^{\frac{3 (n-4)}{8}}
 \xi ^\frac{3 n}{4} {\rho_R}^{3 n/8}}{(n-2)^3  T^\frac{3 n}{2}
   {\rho_S}^{3/2}}.
\end{equation} 
Using \eqref{eq:decaywidth}, the condition for one bubble to be nucleated within each horizon translates now to
\begin{equation}\label{eqn:TnucleationSTIFF}
 \int_{T_n}^{\infty} \Gamma dT=\int_{T_n}^{\infty}\frac{d T}{T}
\frac{M_p^4 2^{\frac{5 n-6}{2}} \left(\frac{3}{\pi }\right)^{\frac{4-n}{2}} \xi ^n {\rho_R}^\frac{n}{2} }{(n-2)^3 T^{2 n-4} {\rho_S}^2}
\exp \left(-\frac{S_3(T)}{T}\right)=1.
\end{equation}

Next, using the Friedmann equation, we can express the new energy density as a function of the ratio of the modified Hubble
 parameter $H$ to the standard radiation-dominated case $H_R$,
as follows 
\be
\rho_S=\left( \left( \frac{H}{H_{R}} \right)^2 -1 \right)
\rho_R 30^{\frac{4-n}{4}} \pi ^{\frac{n-4}{2}} T^{n-4} \left(\frac{{\rho_R}}{g_*}\right)^{\frac{4-n}{4}}
\quad , \quad
\rho_R=\frac{\pi^2}{30}g_* T^4.
\ee
All the above quantities have to be calculated at $T_n$ to match the sphaleron freeze-out calculation, described in the next subsection.

\begin{figure}[h!]
\begin{minipage}[t]{0.45\linewidth} 
\includegraphics[height=6.7cm]{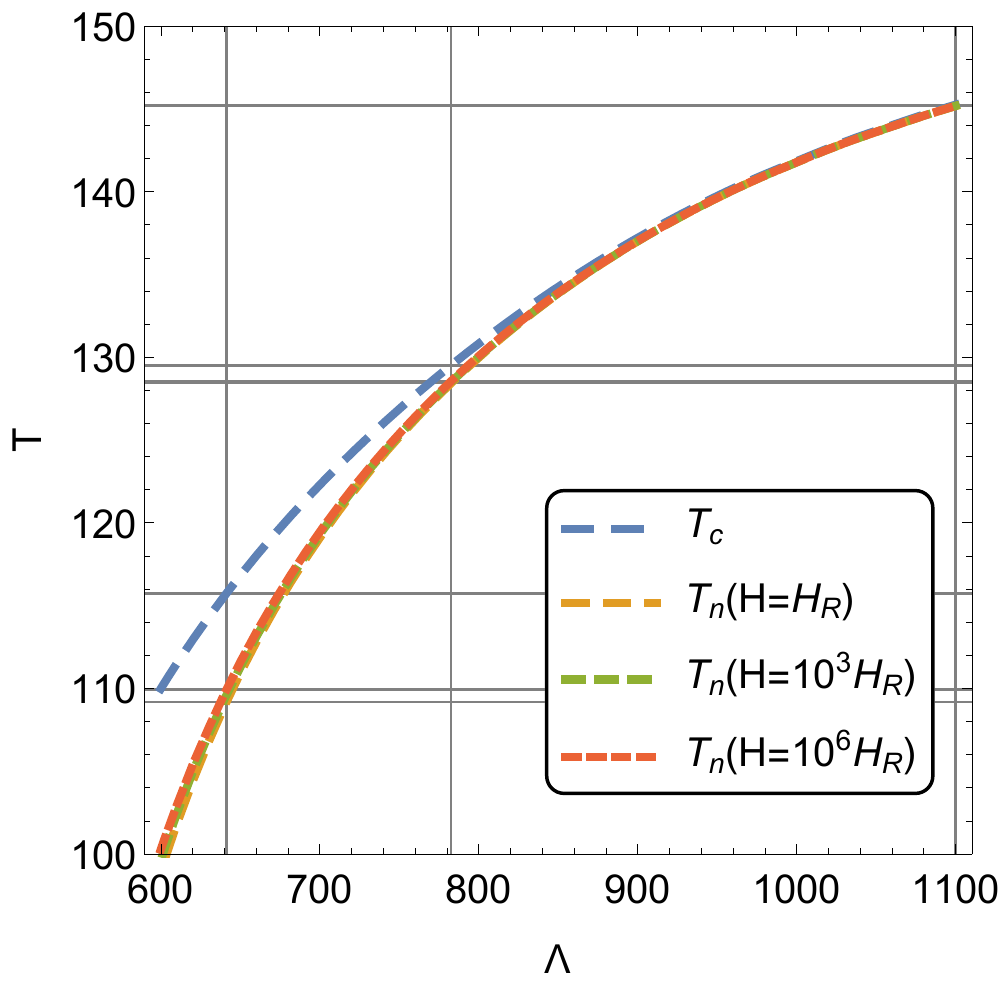} 
\end{minipage}
\begin{minipage}[t]{0.45\linewidth}
\includegraphics[height=6.7cm]{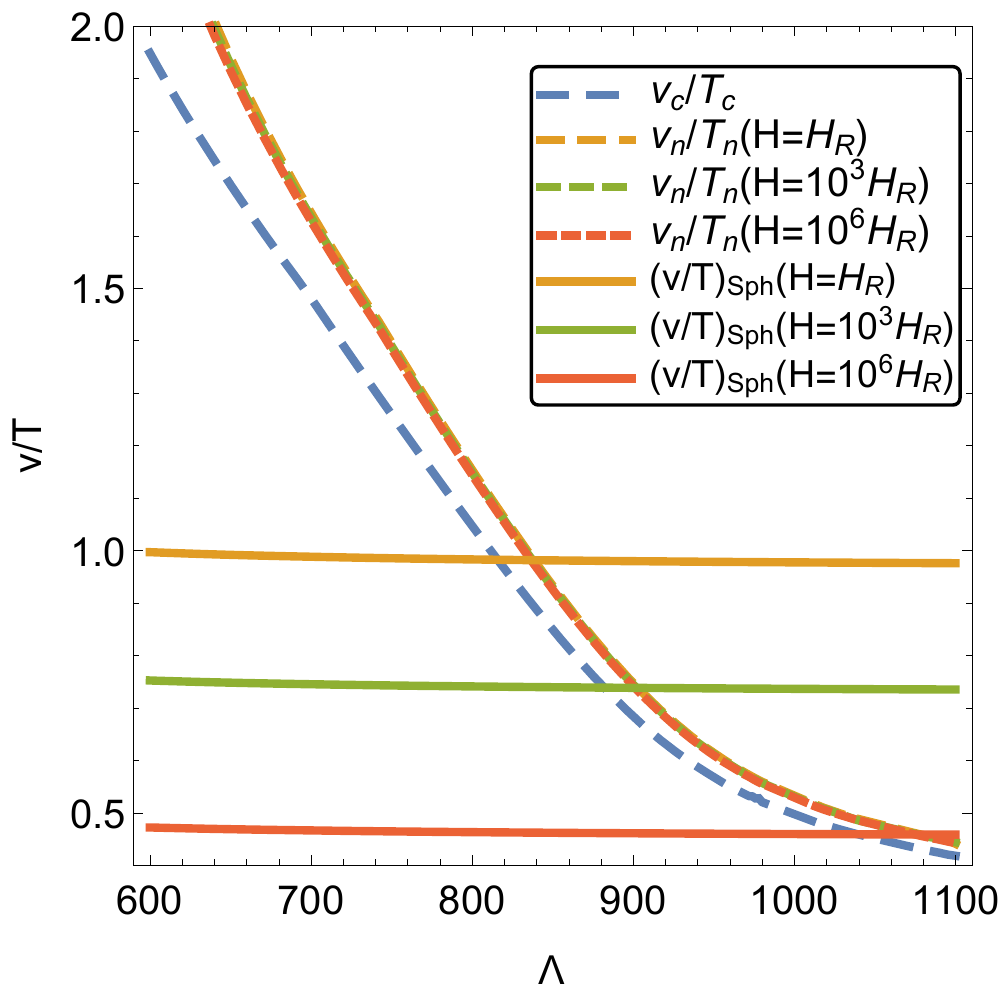}  
\end{minipage}
\caption{
The left panel shows the critical and nucleation temperatures ($T_c $ and $ T_n$) of our model with $n=6$. The thin vertical
lines highlight cutoff scales corresponding to experimental sensitivities, as shown in Figure~\ref{lambda3mplot}, 
while the horizontal lines point to the temperatures corresponding to these values of $\Lambda$. 
The right panel depicts the ratios of vevs to the temperatures ($v(T_c)/T_c $ and $ v(T_n)/T_n$), as a function of $\Lambda$ (dashed lines). 
Also indicated are the sphaleron bounds on $v/T$ for different expansion rates, as described in Sec.\ref{sec:freezout} (solid lines).
\label{Tcplot}
}
\end{figure}

In order to obtain the nucleation temperature, we first solve \eqref{eqn:scalarEOM} numerically using 
an overshoot/undershoot algorithm. This determines the action $S_3(T)$ via \eqref{actionfunc}. 
Finally, we integrate \eqref{eqn:TnucleationSTIFF} to find the nucleation temperature $T_n$ for all values of the cutoff scale $\Lambda$.
Figure~\ref{Tcplot} shows the critical temperature $T_c$ and the nucleation temperature $T_n$,
as well as the ratio of the Higgs vev $v(T):= \langle \phi(T)\rangle$ to those temperatures, $v(T_c)/T_c$ and $v(T_n)/T_n$, as functions of the cutoff scale $\Lambda$,
for the radiation-dominated case ($H=H_R$) and $n=6$ (as for SFDM-domination) with $H=10^3 H_R$ and $H=10^6 H_R$, respectively. 
Also, values of the ratio of the vev to temperature $(v/T)_{\rm Sph}$ required in each of these cases 
by the sphaleron freeze-out (as discussed in Section~\ref{sec:freezout}) are marked in the right panel.

The modification due to the proper calculation of the nucleation temperature can change the resulting bounds 
 significantly, since the difference between these cases increases with the importance of the modification,
which is largest in the most interesting range of small cutoff scales. However,
the correction in \eqref{eqn:TnucleationSTIFF} due to the modified expansion rate only appears in the factor in front of 
the exponential. Therefore, the results are nearly identical for very different ratios of $H/H_{R}$ and $n > 4$.

We can see from the left panel that our EWBG era of interest lies between temperatures of approximately $100 - 150$\gev.
This is the era in which SFDM should be in its stiff phase. 
We presented a fiducial SFDM model in \cite{Rindler-Daller:2015lua}, where we chose a reheating temperature
of $300$ GeV. Reheating is followed by the stiff phase of SFDM, which transitions into its radiation-like phase, making way to radiation-domination before the time of neutron-proton freeze-out. 
The SFDM parameters were $m_s = 5 \times 10^{-21}$ eV and $\lambda_s = 7.5 \times 10^{-59}$ eV cm$^3$ 
(i.e., $\lambda_s/(m_sc^2)^2 = 3 \times 10^{-18}$ eV$^{-1}$ cm$^3$).
Using this SFDM model, the expansion rate during the EWBG era of interest is about $3-5$ orders of magnitudes higher with respect to the
standard case, within the above temperature range.
Work is in progress to study the SFDM parameter space in more detail.

\subsection{Experimental Bounds on Modified Cosmologies}\label{sec:cosmobound}
In this section, we derive very generic bounds which have to be applied to a wide class of cosmological models, and are respected by the SFDM fiducial model as well, 
see \cite{Rindler-Daller:2015lua}.
As elaborated in Sec.\ref{sec:intro} and Sec.\ref{sec:sfdm background}, these bounds come from BBN constraints (see e.g. \cite{Cooke:2014,Agashe:2014kda}), 
whereby current measurements allow for some additional energy density components \cite{Wright:2007vr}.
The simplest way to obtain bounds on this contribution is to translate the bound on the effective number of neutrinos to 
the modification of the Hubble rate \cite{Simha:2008zj},
\begin{equation}
\left. \frac{H}{H_R} \right|_{\rm BBN}=\sqrt{1+\frac{7}{43}\Delta N_{\nu_{\rm eff}}}.
\end{equation}
We will assume $\Delta N_{\nu_{\rm eff}}$ is the difference between the SM radiation contribution $N=3.046$ and the observed value 
$N_{\nu_{\rm eff}}=3.28$ from \cite{Cooke:2014,Agashe:2014kda} which corresponds to $H/H_R|_{\rm BBN}=1.0187$.
It is straightforward to calculate $H_{R,{\rm BBN}}$ using \eqref{eq:rhorad},
\begin{equation}
\rho_{R,{\rm BBN}}=\frac{\pi^2}{30}g_{*, \rm BBN} T_{\rm BBN}^4
\end{equation}
and using the SM values, $T_{\rm BBN}=1$\mev \ and \ $g_{*, \rm BBN}=43/4$.

\begin{figure}[t] 
\begin{center}
\includegraphics[height=6.5cm]{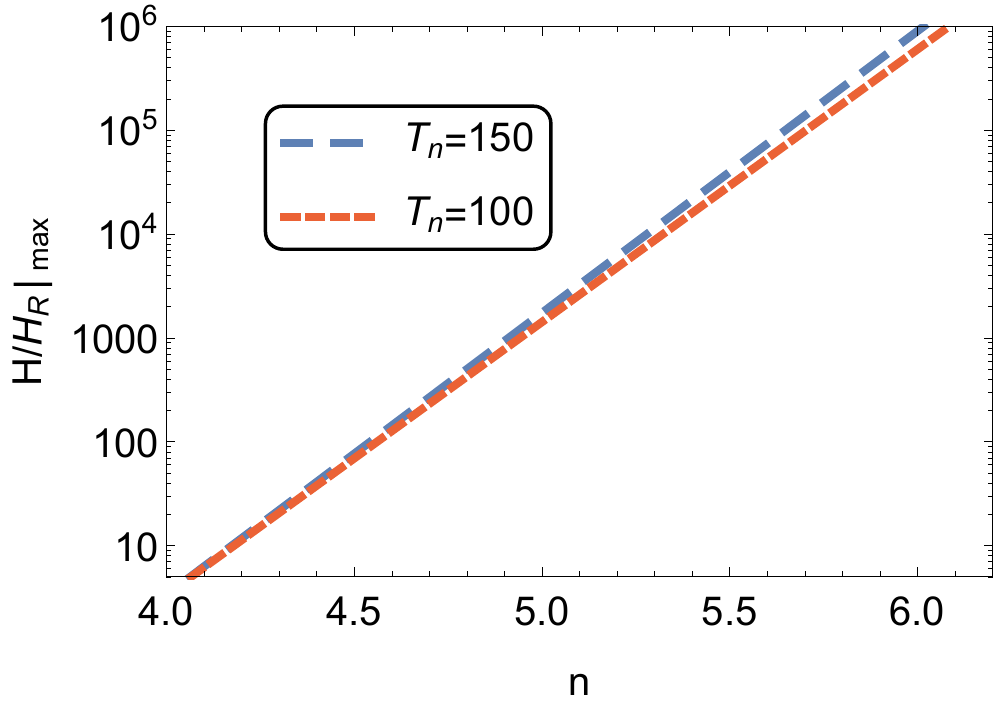} 
\end{center}
\caption{
Maximal modification of the Hubble parameter calculated at the nucleation temperatures $T_n=100$ GeV and $T_n=150$ GeV, as a function of the parameter
 $n$ which determines our cosmological model.
\label{HHRmax}
}
\end{figure}

The next step is to simply compute the energy density of the new component at the EWBG scale, using the Friedmann equation \eqref{eq:SFDMfriedmann}. We assume that 
$\rho_{R, {\rm BBN}}$ is composed of the SM radiation, while the remaining contribution corresponds to
 the new component $\rho_{S,{\rm BBN}}$, 
\begin{equation}
\frac{H}{H_R}=\frac{1}{H_R}\sqrt{\frac{8 \pi}{3 M_p^2}\rho_{S,{\rm BBN}}
\left( \frac{a_{\rm BBN}}{a} \right)^n} =
\sqrt{\left(\left. \frac{H}{H_R} \right|_{\rm BBN} \right)^2-1}
\left(\frac{T_{\rm BBN}}{T_n}\right)^{\frac{4-n}{2}}
\left(\frac{g_{*,{\rm BBN}}}{g_*}\right)^{\frac{1-2n}{4}}.
\end{equation} 
As before, all values without subscript BBN should be calculated at $T_n$. The resulting maximal modification of the expansion rate for different cosmological models in the interesting temperature range $T\in [100,150]$ GeV is shown in Figure~\ref{HHRmax}. 
For our $n=6$ example (i.e., a stiff EOS), this corresponds to a maximal $H/H_R$ ratio between $6\times 10^5$ and $9\times 10^5$, which agrees
 with the results for the fiducial SFDM model, quoted in the previous subsection.
However, the value of $N_{\rm eff} = 3.28$ adopted in this section is smaller than the one chosen for the fiducial model in \cite{Rindler-Daller:2015lua}.  For this $N_{\rm eff}$, the appropriate SFDM model would require a higher mass, while other parameters can stay the same. For instance, for the same ratio of $\lambda_s/(m_sc^2)^2 = 3 \times 10^{-18}$ eV$^{-1}$ cm$^3$ and the same reheating temperature of 300 GeV, SFDM with a mass of 
$m_s = 10^{-20}$ eV would work.

\section{Sphaleron Freeze-Out}\label{sec:freezout}

In this paper, we do not discuss the production of baryons during the phase transition, but rather
the necessary condition for the baryon asymmetry to be not washed out after the phase transition. 
While the $SU(2)$ sphalerons can generate baryon asymmetry \cite{Cohen:1993nk}, they also dilute it
after the phase transition is completed, and the system is returning to thermal equilibrium. 

The sphalerons are suppressed in the broken phase due to the $W$ bosons obtaining mass, and the weak
interactions act only on very short distances. Hence, this suppression is proportional to the gauge boson
masses, thus proportional to the Higgs vev right after the phase transition. If the barrier separating the vacua is 
too thin and the broken phase vev is too small, all the generated asymmetry can be washed out, including asymmetry 
which may have been generated in mechanisms different from EWBG, as long as these mechanisms
also respect the $B-L$ 
symmetry of the SM \cite{Quiros:1999jp}.  

\begin{figure}[t] 
\begin{minipage}[t]{0.45\linewidth} 
\includegraphics[height=6.7cm]{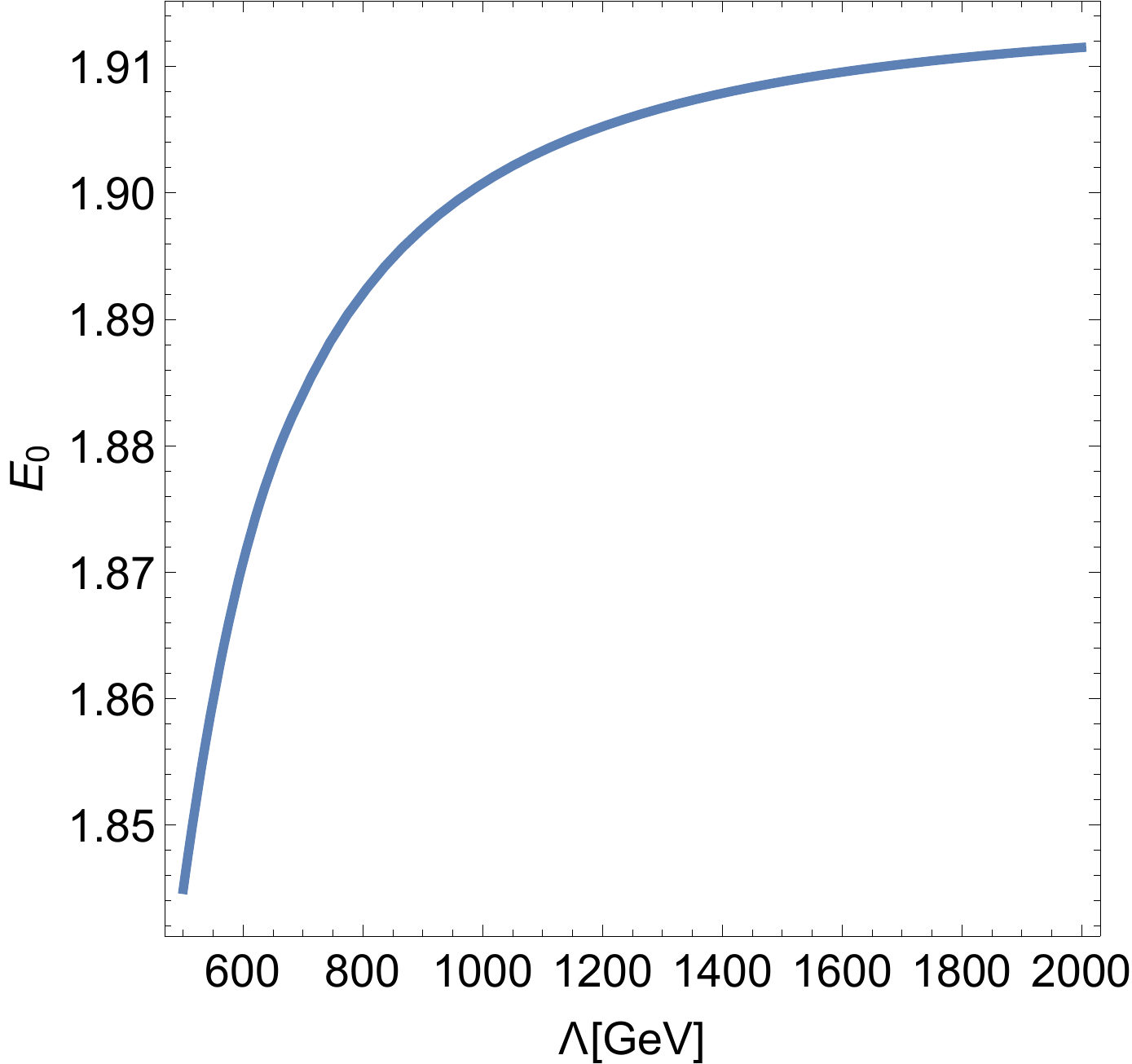} 
\end{minipage}
\begin{minipage}[t]{0.45\linewidth} 
\includegraphics[height=6.7cm]{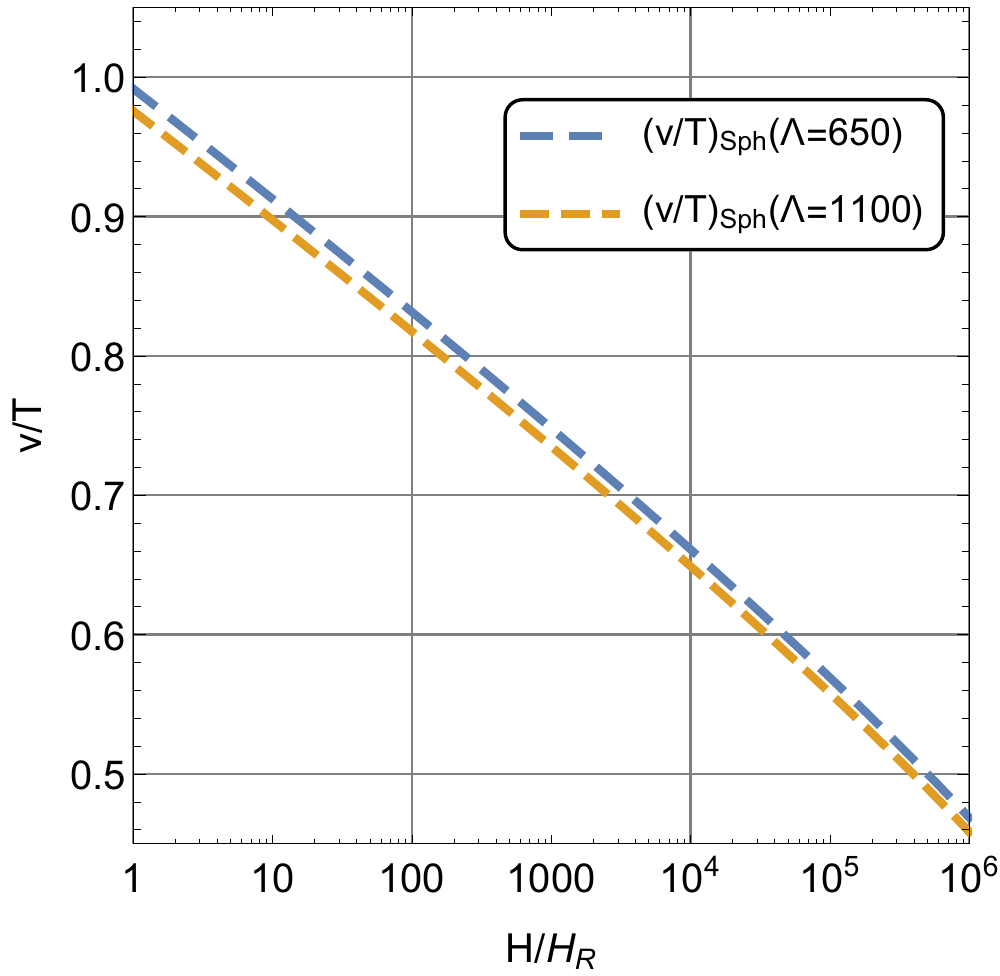}
\end{minipage}
\caption{
Left panel: sphaleron energy (divided by the factor of $4 \pi v /g$) as a function of the cutoff scale $\Lambda$.
Right panel: minimal value of $v/T$ required to avoid wash-out of baryon asymmetry, as a function of the modified expansion rate
 $H/H_R$ for several values of the cutoff $\Lambda$.
\label{E0plot}
}
\end{figure}

A simple criterion for sphaleron freeze-out (or ``wash-out'') is obtained by assuming that the sphaleron processes decouple
when their rate becomes smaller than the expansion rate of the Universe, i.e., when $\Gamma \lessapprox H$.
The sphaleron rate is given by \cite{Quiros:1999jp}
\be
\Gamma = 2.8\times 10^5 T^4 \kappa \frac{g}{4 \pi}\left(\frac{v}{T}\right)^7 \exp \left( -\frac{E_{\rm sph}}{T} \right),
\ee
where the parameter $\kappa$ is the functional determinant associated with fluctuations about the sphaleron.
That parameter has been estimated to be in the range $10^{-4} \lesssim \kappa \lesssim 10^{-1}$.
The sphaleron energy $E_{\rm sph}$ is modified due to the cutoff $\Lambda$ \cite{Grojean:2004xa}, as well as due 
to the exponential dependence on the action. Indeed, this can have a significant impact on the bounds we can place on $\Lambda$. 
In order to accurately calculate the sphaleron energy
we find the sphaleron solution by starting with the 
ansatz~\cite{Klinkhamer:1984di} for the $SU(2)$ gauge field $W$ and the scalar field $\phi$, 
\begin{eqnarray}
\nonumber
W^a_i\sigma^adx^i   =  -\frac{2i}{g} f(\xi) dU\, U^{-1}, \ \
\phi  =  \frac{v_0}{\sqrt{2}} h(\xi) U
\left(\begin{array}{c} 0 \\  1\end{array}\right), 
\nonumber
\end{eqnarray}
where $\xi=gv_0r$, $\sigma^a$ are the Pauli matrices and $\displaystyle U =\frac{1}{r}\left( 
\begin{array}{cc}
z & x+iy \\
-x+iy & z 
\end{array}
\right)$, while $f$ and $h$ are unknown functions of the single variable $\xi$. 
We will compute only the $SU(2)$ sphaleron,
neglecting small corrections from $U(1)_Y$, as in \cite{Klinkhamer:1984di}.
With the above assumptions, the action of the sphaleron reads $E_{sph}=(4\pi v/g)E_0$ with
{\small
\be\label{eq:sphaleron_energy} 
E_0 = \int_0^\infty d\xi \left( 4f'^2
+\frac{8}{\xi^2}f^2(1-f)^2+\frac{1}{2}\xi^2 h'^2 +\right.
 h^2(1-f)^2+\frac{\lambda}{4g^2}\xi^2 (h^2-1)^2  +\left.\frac{v^2}{8g^2\Lambda^2}\xi^2 (h^2-1)^3\right).
\ee}
Varying this action, we find the field equations for the functions $f$ and $h$,
\begin{eqnarray}\label{eq:sphal_eom}
\xi^2 \frac{d^2f}{d\xi^2}& = & 2f(1-f)(1-2f)-\frac{\xi^2}{4}h^2(1-f)  \\
\frac{d}{d\xi}\left[ \xi^2\frac{dh}{d\xi}\right]& = & 2h(1-f)^2 
 +\frac{\lambda}{g^2}\xi^2(h^2-1)h +\frac{3}{4}\frac{v_0^2}{g^2\Lambda^2}\xi^2 h(h^2-1)^2.
\nonumber
\end{eqnarray}
These are subject to the boundary conditions $f(0)=h(0)=0$ and $f(\infty)=h(\infty)=1$.
In order to find the exact solutions, we start with the analytical solutions of the asymptotic equations, valid 
near the boundaries as shown in Table~\ref{tab:asymptoticsol}.
\begin{table}[t]
\centering
\begin{tabular}{c | c}
$\xi \rightarrow 0$ &  $\xi \rightarrow \infty$  \\
\hline
\vspace{-0.4cm}
\\
$f \approx  \xi^2/a_0^2$ & $f \approx  1-a_\infty\exp (-\xi /2) $\\
$h \approx \xi/b_0$ &$h \approx  1-(b_\infty/\xi)\exp (-\sqrt{\frac{2\lambda}{g^2}} \xi)$
\end{tabular}
\caption{Analytic solutions of the asymptotic equations of motion (\ref{eq:sphal_eom}) describing
the sphaleron solution near the boundaries.}
\label{tab:asymptoticsol}
\end{table}
Using these solutions to find our initial conditions at a certain very small and very large value of $\xi$,
we numerically solve the full equations 
to a certain $\xi_{match}$ where we compare the two solutions. 
Our procedure consists of randomly varying the initial parameters $a_0$, $b_0$, $a_\infty$ and $b_\infty$
and updating them if solutions with the new values match more closely. When both functions and their derivatives
at $\xi_{match}$ match with an accuracy of $10^{-6}$, we consider the equations solved, and use that solution to 
calculate the resulting sphaleron energy, $E_{\rm sph}= (4\pi v/g) E_0$ from \eqref{eq:sphaleron_energy}. 
Now, we can rewrite the freeze-out condition $\Gamma \leq H$ as
\begin{equation}\label{eq:requiredvoverT}
\frac{v}{T} \geq \frac{g}{4 \pi E_0}\ln \left( \frac{ 2.8\times 10^5  T^4 \kappa \frac{g}{4 \pi}\left(\frac{v}{T}\right)^7 }{H} \right),
\end{equation}
where $H$ is the Hubble rate calculated at the nucleation temperature $T_n$ when the phase transition ends. We will choose 
$\kappa =10^{-1}$, which gives the most stringent constraints.
The (weak) dependence of $E_0$ on $\Lambda$ is shown in the left panel of Figure~\ref{E0plot}. It is also the reason why the $(v/T)_{Sph}$-lines shown 
in Figure~\ref{Tcplot} are not straight, but show a slight dependence on $\Lambda$. We can see that the sphaleron 
energy decreases for smaller cutoffs. This means that the sphaleron processes
are more active, hence more suppression is required after the phase transition. However, the effect is rather weak and changes
the results only by a small amount, compared to the changes coming from the modified expansion history,
as can be seen in the right panel of Figure~\ref{E0plot}. 

\begin{figure}[t] 
\begin{center}
\includegraphics[height=6.5cm]{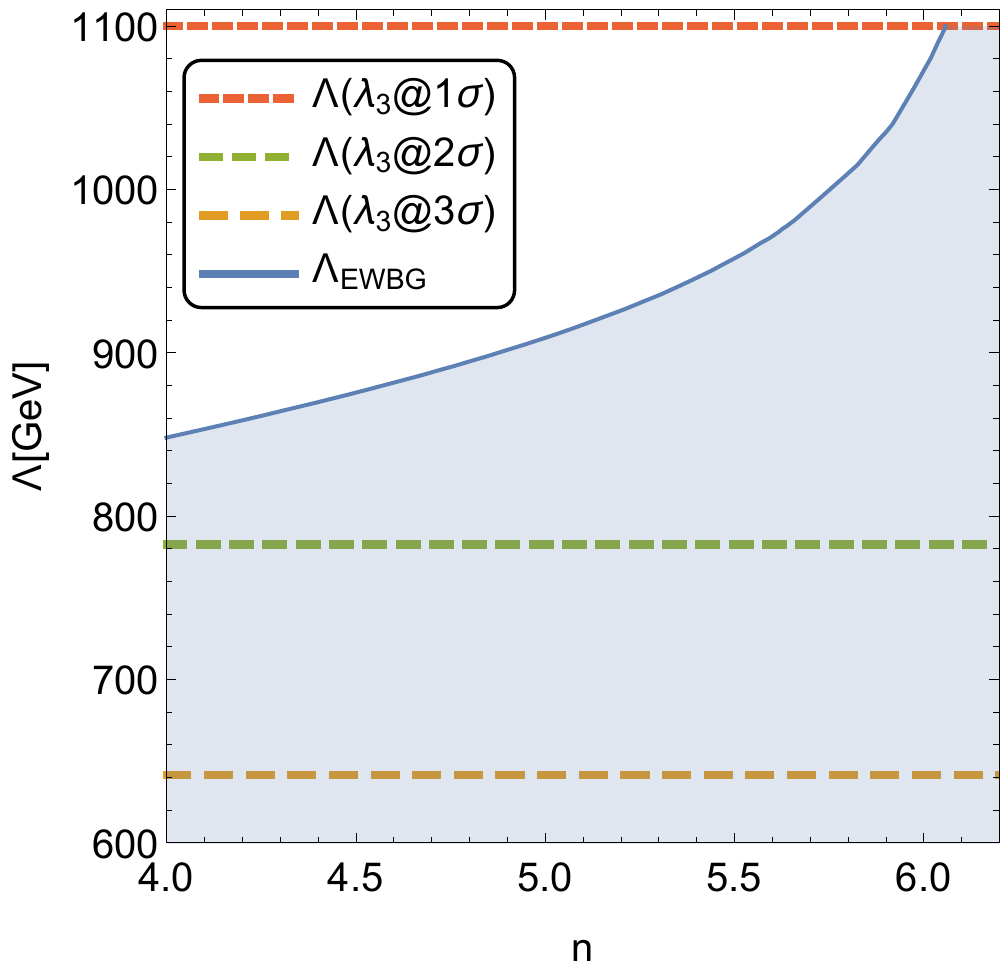} 
\includegraphics[height=6.5cm]{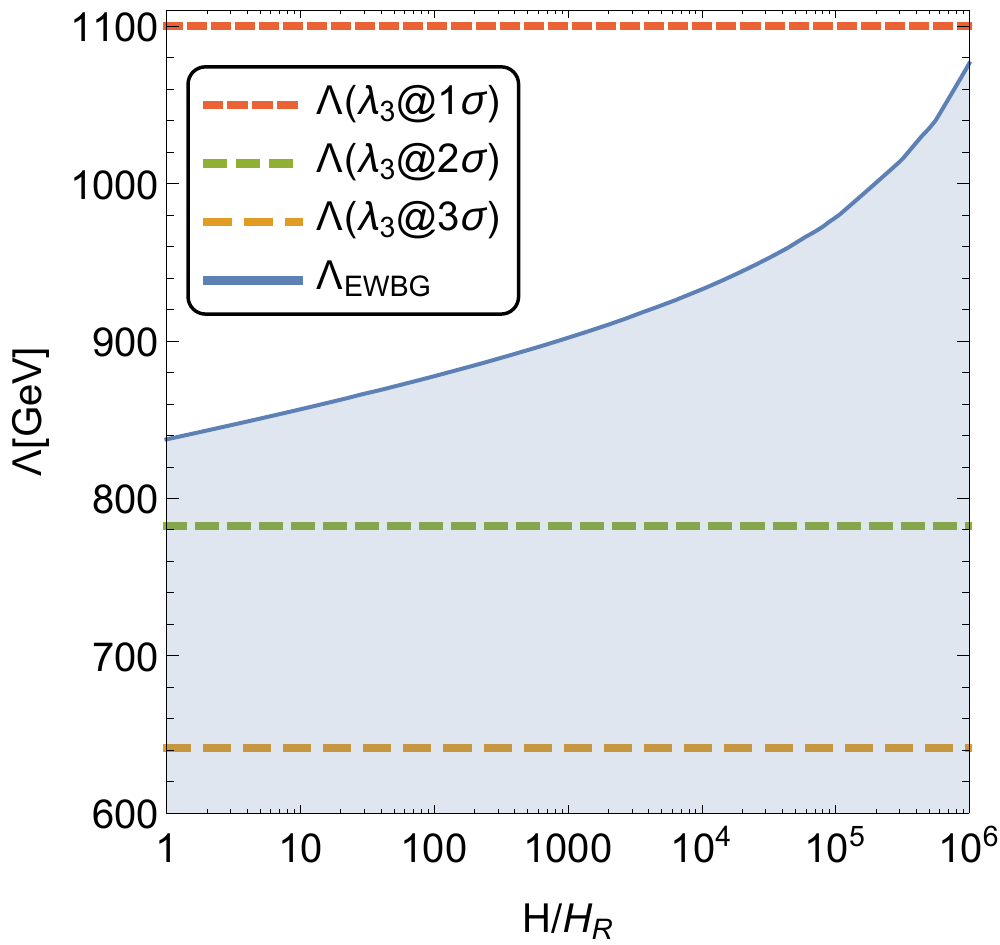} 
\end{center}
\caption{
Left panel: cutoff scale $\Lambda$ required for successful EWBG  (solid blue line) as a function of $n$ which determines our cosmological model. Here we assumed maximal experimentally allowed assistance from cosmology. 
Right panel: cutoff scale $\Lambda$ required by EWBG as a function of the expansion rate (solid blue curve) for $n=6$.
Both panels also show HL-LHC experimental constraints on $\Lambda$ from its modification to $\lambda_3$ (horizontal dashed lines, $1\sigma$ (top) to $3\sigma$ (bottom)).
\label{finalplot}
}
\end{figure}

We are finally in position to combine the $v/T$ value required to decouple the sphalerons and preserve the 
asymmetry \eqref{eq:requiredvoverT} with the $v/T$ value we obtain as a function of the cutoff from Figure~\ref{Tcplot},
along with the experimental constraints on the cutoff from Figure~\ref{lambda3mplot}. 
Thus, we can determine the minimal scale of new physics required to preserve the asymmetry as a function of 
the modified expansion rate $H/H_R$. 
We can also translate the maximal possible modification of the Hubble rate, discussed in Section~\ref{sec:cosmobound}, to an explicit bound on 
$\Lambda$ for a wide class of cosmological models.
Figure~\ref{finalplot} shows the minimal value of $\Lambda$ as a function of $n$, along with the experimental constraints, and the specific cosmological example for $n=6$ 
(i.e., a stiff EOS, as in SFDM).

Our key result is that for $n=6$,  the minimal $\Lambda$ required by the sphaleron bound is already very close to the value required for the first order phase transition (as discussed in Section~\ref{sec:particlemodel}).
Thus, the modified cosmological history allows us to circumvent the sphaleron bound altogether, and the only bound given by current experiments is equivalent to the requirement of departure from thermal equilibrium. 


%
%

\section{Conclusions}\label{sec:conclude}

In this paper we studied the implications of a modified cosmological history for the electroweak baryogenesis scenario.
We adopted a generic model in which the Higgs potential is modified by a non-renormalizable dimension six operator, suppressed by an appropriate new mass scale $\Lambda$.

We discussed a very generic model of cosmological modification with a single new energy density component which does not interact with SM degrees of freedom. 
As a possible specific source of such a modification, we focused on complex scalar field dark matter (SFDM).

We carefully computed the temperature at which the phase transition takes place, instead of using the approximation coming from the critical temperature, often used in the literature. 
This allowed us in addition to include minor corrections to the nucleation temperature due to a modification of the cosmological history. In all, using the nucleation temperature in the full calculation, rather than the critical temperature approximation, can change the final results significantly for the allowed parameter space.

Next, we described the modification of Standard Model $SU(2)$ sphalerons. 
This is the main source of modification resulting from the increased expansion rate.
A higher expansion rate leads to a more readily achieved freeze-out of the sphalerons, thus preserving any baryons remaining after the phase transition.
This in turn increases the minimal scale $\Lambda$ of new physics which is required for successful baryogenesis.

We find that this modification of the required $\Lambda$'s, while numerically seemingly small (about 20\% for $\rho_S\propto a^{-6}$), actually means circumventing the sphaleron bound altogether, because it brings us to the cutoff values required for a first order phase 
transition to begin with. 
Also, our specific example of SFDM, and other models with $n=6$, prove very interesting, since they allow us to get very close to avoiding the sphaleron bound.
Exotic models with even higher expansion rates (i.e., whose energy density would decay even faster than $\propto a^{-6}$) would not increase the allowed parameter space much further. 
On the particle experimental side, it means that, with assistance of a fluid with stiff EOS, like SFDM, our model can remain consistent within $1 \sigma$ of the
SM result, even with the bounds provided by the high luminosity stage of the LHC. 


\section*{Acknowledgements}
We would like to thank M. Artymowski for very helpful discussions.
ML was supported by the Polish National Science Centre under research grant 2014/13/N/ST2/02712 and 
doctoral scholarship number 2015/16/T/ST2/00527. TRD was supported by the U.S. Department of Energy
under grants DE-FG02-95ER40899 and DE-SC0007859. JW is supported in part by the Department of Energy under grant DE-SC0007859.

\bibliographystyle{JHEP}
\bibliography{EWBGbibliography}
\end{document}